\documentclass[pss,fleqn]{w-art}
\usepackage{times}
\usepackage{w-thm}
\usepackage{graphicx}
\usepackage{cite}
\usepackage{amssymb}
\chardef\bslash=`\\ 

\begin{document}
\DOIsuffix{theDOIsuffix}
\Volume{XX}
\Issue{1}
\Month{01}
\Year{2003}
\pagespan{1}{}
\Receiveddate{XX April 2004}
\Reviseddate{XX April 2004}
\Accepteddate{XX April 2004}
\Dateposted{$Revision: 1.13 $, compiled \today}
\keywords{off-diagonal disorder, two-dimensional systems, localization,
  finite-size scaling, critical exponents}
\subjclass[pacs]{72.15.Rn, 73.30.Fz}

\title[Exponents of the localization length in 2D]{Exponents of the
  localization length in the 2D Anderson model with off-diagonal
  disorder}

\author[Eilmes]{Andrzej Eilmes\inst{1}}

\address[\inst{1}]{Department of Computational Methods in Chemistry,
  Jagiellonian University, Ingardena 3, 30-060 Krak\'{o}w, Poland}

\author[R\"{o}mer]{Rudolf A.\ R\"{o}mer\footnote{Corresponding author:
    e-mail: {\sf r.roemer@warwick.ac.uk}}\inst{2)}}

\address[\inst{2}]{Department of Physics and Centre for Scientific
  Computing, University of Warwick, Coventry, CV4 7AL, United Kingdom}

\begin{abstract}
  We study Anderson localization in two-dimensional systems with purely
  off-diagonal disorder. Localization lengths are computed by the
  transfer-matrix method and their finite-size and scaling properties
  are investigated.  We find various numerically challenging differences
  to the usual problems with diagonal disorder. In particular, the
  divergence of the localization lengths close to the band centre is
  investigated in detail for bipartite and non-bipartite lattices as
  well as different distributions of the off-diagonal disorder.
  Divergence exponents for the localization lengths are constructed that
  appear to describe the data well down to at least $10^{-5}$. We find
  only little evidence for a crossover energy scale below which the
  power law has been argued to fail.
\end{abstract}

\maketitle

\section{Introduction}
\label{sec-intro} 

The disorder-induced metal-insulator transition and the concept of
Anderson localization \cite{And58,EcoC70,EcoC72,Eco72,LicE74} have been
the subject of an intense research for more than forty years. The highly
successful scaling approach for non-interacting electrons was proposed
by Abrahams {\it et al.} \cite{AbrALR79} in 1979, predicting that for
generic situations in 1D and 2D all states remain localized, thus there
is no disorder-driven transition \cite{LeeR85,KraM93,BelK94}.
However, it was suggested \cite{EcoA77,AntE77,Weg79} that an Anderson
model with purely off-diagonal disorder may be an exception, as
non-localized states were found at the band center
\cite{Oda80,SouE81,FerAE81,PurO81,SouWGE82}. Recent numerical
investigations \cite{EilRS98a,EilRS98b,Cai98T,CaiRS99,BisCRS99,XioE01}
have revealed unusual localization properties of these states.  It was
found that the localization length diverges at the energy $E=0$
\cite{EilRS98a,EilRS98b}, scaling properties of this divergence suggest
that the states at the band center are critical \cite{EilRS98b,XioE01},
thus there are no truly extended states in 2D in agreement with the
scaling arguments.

Of special interest is the model of off-diagonal disorder on the
bipartite lattice with even number of sites, which exhibits special
symmetry properties at the band center --- the energy spectrum is
symmetric around $E=0$. It has been shown that in this case, related to
the chiral universality class \cite{BocC03,MudRF03}, states at the band
center are non-localized in any dimension \cite{Weg79,GadW91,Gad93}.
This has been recently demonstrated in 2D and 3D using a renormalization
group (RG) approach \cite{FabC00}.
In a recent paper \cite{EilRS01}, we studied 2D bipartite systems with
various types of off-diagonal disorder by means of the transfer-matrix
method (TMM) and investigated the divergence of the localization lengths
close to the band center. In Ref.\ \cite{FabC00} it has been suggested
that this divergence may be described by
\begin{equation}\label{eq-power-law}
   \xi(E) \propto \left| \frac{E_0}{E} \right|^{\nu}, \quad |E| > E^{*}, 
\quad\hfil\mbox{and}\hfil\quad
   \xi(E) \propto \exp\sqrt{\frac{\ln E_0/E}{A}}, \quad |E| < E^{*}
\end{equation}
with a certain, unspecified crossover energy $E^{*}$ distuinguishing
between the power law and the more complicated form. In \cite{EilRS01}
we showed that a power-law behavior fits the data down to energy $E = 2
\times 10^{-5}$ and the corresponding divergence exponents $\nu$ have
rather low values $\sim 0.3$ that seem to depend non-universally on the
type and strength of the disorder.
In this paper we extend these calculations to energies even closer to
the band center at $E=0$. In addition to the square 2D lattices we
examine also honeycomb lattices where the Van Hove singularity at $E=0$
does not interfere with the divergence due to the chiral symmetry.

\section{The Hamiltonian, off-diagonal disorder distributions and 
  the transfer-matrix method}
\label{sec-model}

The Anderson Hamiltonian for a single electron on a 2D lattice is
\begin{equation}
  H = \sum_{i \neq j}^N t_{ij} \left| i \right\rangle \left\langle j
  \right| + \sum_i^N \epsilon_i \left| i \right\rangle \left\langle i
  \right|
\label{eq-hamilt}
\end{equation}
where $\left| i \right\rangle$ denotes the electron wave function at
site $i$. For purely off-diagonal disorder, we set the onsite energies
$\epsilon_i$ to $0$. The off-diagonal disorder is introduced by random
hopping elements $t_{ij}$ between nearest neighbor sites. In a square
lattice all four hopping elements to the four nearest neighbor are
chosen according to the random distribution. The honeycomb lattice
\cite{SchO91} is topologically equivalent to the brick-layer structure
\cite{GriRS98} --- the corresponding hopping element of the underlying
square lattice is equal $0$.

On both lattices we study three different distributions of off-diagonal
elements $t_{ij}$,
\begin{eqnarray}
P\left(t_{ij}\right) & =
\left\{ \begin{array}{ll}
    1/w & \textrm{if $\left|t_{ij}-c\right| \leq w/2$,}\\
    0   & \textrm{otherwise,}
\end{array}\right.
& \mbox{a box distribution \cite{EilRS98a}},\\
 P\left(t_{ij}\right) & = \frac{1}{\sqrt{2\pi\sigma^{2}}}
            \exp \left[ -\frac{\left(t_{ij}-c\right)^2}{2\sigma
    ^2} \right],
& \mbox{a Gaussian distribution},\\
P\left(\ln t_{ij}/t_0\right) & = \left\{ \begin{array}{ll}
       1/w & \textrm{if $\left| \ln t_{ij}/t_0 \right| \leq w/2$,}\\
       0   & \textrm{otherwise.}
        \end{array} \right.
& \mbox{a box distribution for $\ln t_{ij}$ \cite{SouWGE82}}.
\end{eqnarray}
The logarithmic $t$ ($\ln t$) distribution is believed to be more
suitable to model actual physical systems \cite{SouWGE82} and the
parameter $w$ is a good measure for the off-diagonal disorder strength.
We will show however, that although it directly avoids problems with
zero $t$ elements present in the case of box and Gaussian distributions
\cite{EilRS98a}, for larger $w$ values it is likely to suffer similar
numerical problems due to a large number of small elements close to
$\exp(-w)$.

The TMM \cite{MacK81,MacK83} is widely used to study the localization
properties of states in disordered systems.  To calculate the decay
lengths of wave functions on quasi-1D strips of width $M$ and length $K
\gg M$ the Schr\"{o}dinger equation for the Hamiltonian (\ref{eq-hamilt})
is written in the TMM form:
\begin{equation}
\label{eq-tmm}
\left(
\begin{array}{c}
\psi _{n+1}\\
\psi _n
\end{array}
\right) =
 \left(
\begin{array}{cc}
[t^{||}_{n+1}]^{-1}(E-\epsilon_n - H_{\perp}) & - [t^{||}_{n+1}]^{-1}t^{||}_{n} \\
1 & 0
\end{array}
\right)
\left(
\begin{array}{c}
\psi _n \\
\psi _{n-1}
\end{array}
\right) =
 T_n\left(
\begin{array}{c}
\psi _n \\
\psi _{n-1}
\end{array}
\right),
\end{equation}
where $\psi_n= (\psi_{n,1}, \psi_{n,2}, \ldots, \psi_{n,M})^T$ denotes
the wave function at all sites of the $n$th slice, $\epsilon_n= {\rm
  diag}(\epsilon_{n,1}, \ldots, \epsilon_{n,M})$, $H_{\perp}$ is the
hopping Hamiltonian within slice $n$ and $t^{||}_n=$ ${\rm
  diag}(t^{||}_{n,1}$, $t^{||}_{n,2}$, $\ldots$, $t^{||}_{n,M})$ is the
diagonal matrix of hopping elements connecting slice $n-1$ with slice
$n$.
For rectangular and Gaussian distributions of hopping elements we set
the width $w$ and the standard deviation $\sigma$ of the distribution to
$1$ and center it at the $c=0$. In the case of the $\ln t$ distribution
we chose $t_0 = 1$ which sets the energy scale and perform calculations
for two values $w=2$ and $6$ of the distribution width.

In the case of the honeycomb lattice half of the connections to the
nearest neighbors perpendicular to the TMM direction are missing, thus
the lattice topology is reflected in the $H_{\perp}$ part of the
Hamiltonian.  The largest localization length
$\lambda(E,M)=1/\gamma_{\rm min}$ at energy $E$ and strip width $M$ is
determined by the smallest Lyapunov exponent $\gamma_{\rm min}>0$
obtained as the eigenvalue closest to one of the product of the transfer
matrices $\tau_K = T_K T_{K-1} \ldots T_2 T_1$ when $K\rightarrow
\infty$ \cite{Ose68}.

After calculating localization lengths $\lambda(M)$ for increasing
widths of the strips $M$ we scaled the reduced localization lengths
$\lambda(M)/M$ onto a single scaling curve
\begin{equation}
\label{sc-curv}
\lambda (M)/M = f(\xi/M).
\end{equation}
The scaling function and the values of the scaling parameter $\xi$ were
determined by the finite-size scaling (FSS) procedure as in Ref.\ 
\cite{MacK83}. As we will explain below, the recent FSS approach as
outlined in Refs.\ \cite{SleO99a,MilRSU00} in not applicable here since
the functional form of the divergence of $\xi(E)$ for $E\rightarrow 0$
is not yet clear.

\section{Singularities in the density of states}
\label{sec-dos}

The square lattice without disorder exhibits a van-Hove singularity in
the density of states (DOS) at the band center as shown in Fig.\ 
\ref{fig-DOS}. Its properties could mask the Dyson singularity due to
the off-diagonal disorder. Thus we also investigate a honeycomb lattice
where the van-Hove singularity is absent at $E=0$. In fact the DOS for
the ordered system on honeycomb lattices exhibit a dip at $E=0$
\cite{GriRS98}. With increasing off-diagonal disorder the DOS at this
energy increases and for large disorders it becomes as large as for
square lattices. These features are readily visible in Fig.\ 
\ref{fig-DOS}.
\begin{vchfigure}[htb]
\centerline{\includegraphics[width=.95\textwidth]{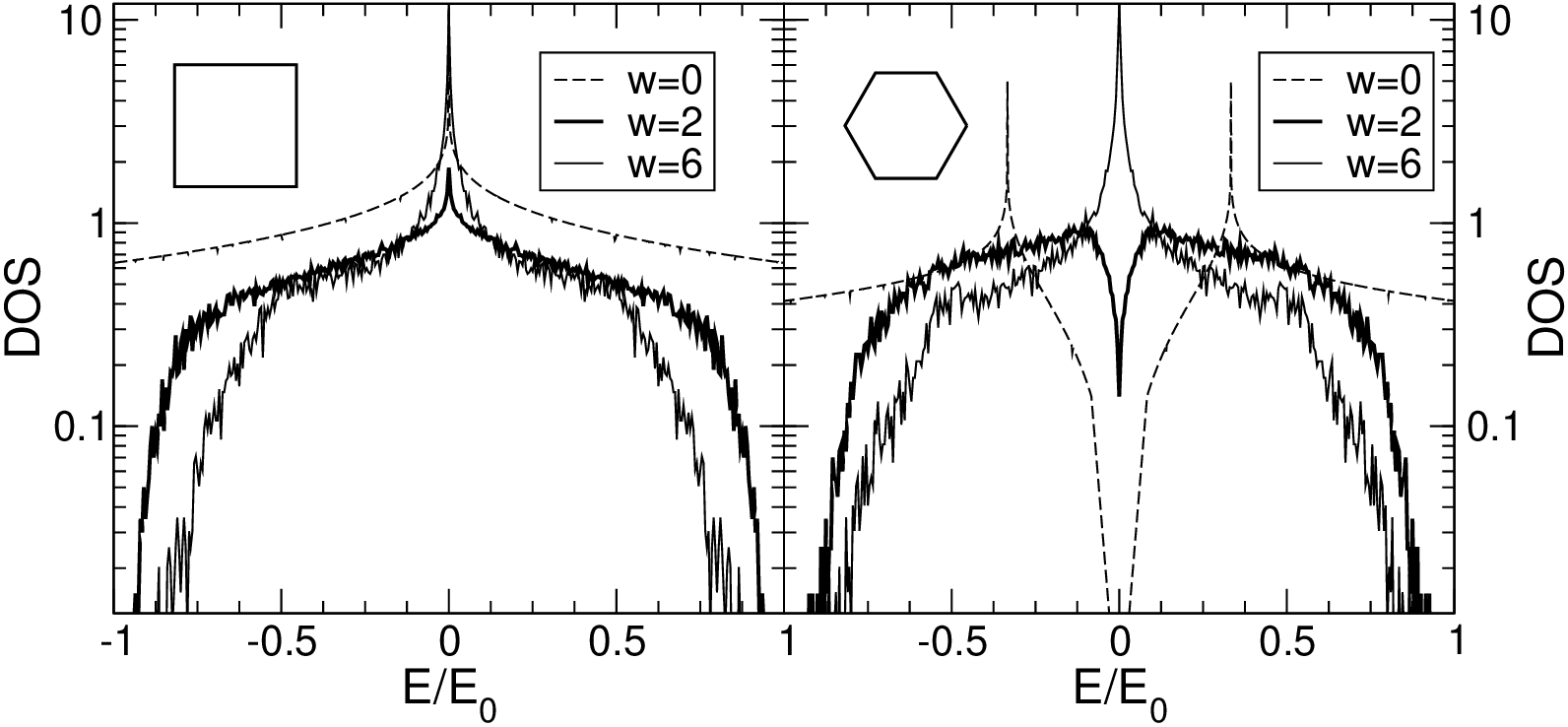}}
\caption{
  Density of states for purely off-diagonal disorder with a $\ln t$
  distribution on a square (left) and a honeycomb (right) lattice. The
  data has been obtained by exact diagonalization of the model
  (\protect\ref{eq-hamilt}) for $200 \times 200$ sites for one
  realization of the disorder. The dashed lines correspond to the
  ordered lattices. In all cases, the energy has been rescaled by the
  band width and the DOS has not been normalized to integrate to unity.}
\label{fig-DOS}
\end{vchfigure}
It is worth noting, that for large $w$ values the peak at the band
center for the $\ln t$ distribution is much higher than in case of
rectangular or Gaussian distribution.

\section{Localization properties of bipartite lattices}
\label{sec-bipartite}

The TMM calculations are performed for strip widths up to $M=220$ in the
energy interval $1 \times 10^{-8} \leq E \leq 0.1$, the actual values
depend on the disorder parameters. The accuracy of the localization
lengths is $1\%$ in all cases.  

\subsection{Finite-size effects and scaling close to the band center}
\label{sec-calc}

In Fig.\ \ref{fig-evodd}, we show the behavior of the localization
lengths for various strip widths as a function of energy. The
$\lambda_M/M$ data for small system sizes shows a pronounced
bending-down effect close to $E=0$ which is most prominent for even
strip widths \cite{BisCRS99}. For odd strip widths $\lambda_{M}/M$
monotonically increases (in the examined energy range) as the energy
approaches $0$, although for the smallest width it is almost constant
close to the band center (\ref{fig-evodd}, right panel).
\begin{vchfigure}[thb]
\centerline{\includegraphics[width=.95\textwidth]{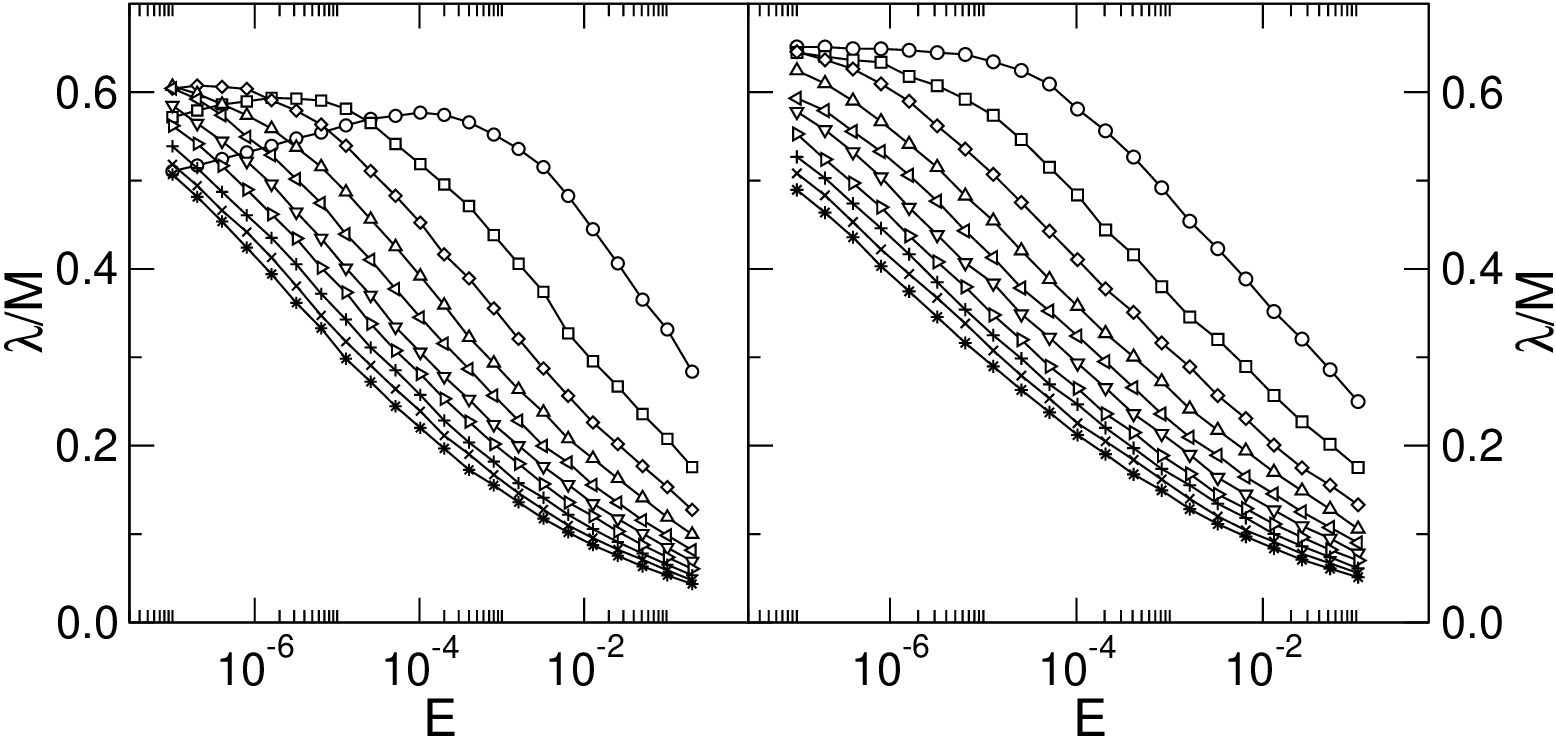}}
\caption{
  Reduced localization lengths $\lambda_M/M$ on a square lattice as
  function of energy $E$ at disorder $w=6$. Left: {\em even} system
  widths ranging from $M= 10$ ($\circ$), $20$ ($\Box$), to $100$ ($*$);
  Right: {\em odd} widths from $M= 15$ ($\circ$), $25$ ($\Box$), to
  $105$ ($*$).  }
\label{fig-evodd}
\end{vchfigure}

Clearly, such a behavior at small $M$ can not be captured easily in an
FSS procedure. Therefore we only use system widths large enough such
that the bending for small energies is irrelevant. Of course this
increases the computational effort necessary. Furthermore, the
accessible strip widths will depend on the value of the disorder
parameter and energy, smaller values requiring much more time. In Table
\ref{tab-res} we show the list of values used.
\begin{table}[htb]
\caption{
  Estimated values of the exponents of the localization lengths defined 
  in (\protect\ref{eq-power-law}) for  
  various disorder strengths and distributions. The error bars only 
  represent the standard deviations from a power-law fit of 
  (\protect\ref{eq-power-law}) to the $\lambda(E,w)$ data and should
  be increased up to one order of magnitude for a reliable
  representation of the actual errors.}
\label{tab-res}
\renewcommand{\arraystretch}{1.5}
\begin{center}
\begin{tabular}{l|c|c|c|c}
 &  \multicolumn{2}{c}  {square lattice}   \vline
 &  \multicolumn{2}{c} {honeycomb lattice}
\\
\hline
disorder parameters  & sizes used in FSS  & $\nu$ &
sizes used in FSS & $\nu$\\
\hline
box, $c=0$      & $150-220$ & $0.317 \pm 0.007$ & $110-170$ & $0.290 \pm 0.004$ \\
Gaussian, $c=0$ & $110-160$ & $0.303 \pm 0.006$ & $120-170$ & $0.273 \pm 0.005$ \\
$\ln t$, $w=2$      & $110-190$ & $0.357 \pm 0.009$ & $100-170$ & $0.604 \pm 0.015$ \\
$\ln t$, $w=6$      & $120-170$ & $0.232 \pm 0.007$ & $120-170$ & $0.238 \pm 0.007$ \\
\end{tabular}
\end{center}
\end{table}

Next, an FSS procedure \cite{MacK83} unbiased by any preset fitting
function is applied to the data and the infinite-size localization
length $\xi$ is extracted.  Fig.\ \ref{fig-fss-sq}  shows the resulting 
scaling plots. We emphasize that this FSS procedure does not require 
any apriori knowledge of which function in (\ref{eq-power-law}) is correct 
and does not make any assumptions in the general form of the scaling 
function $f$.
\begin{vchfigure}[htb]
\centerline{\includegraphics[width=.95\textwidth]{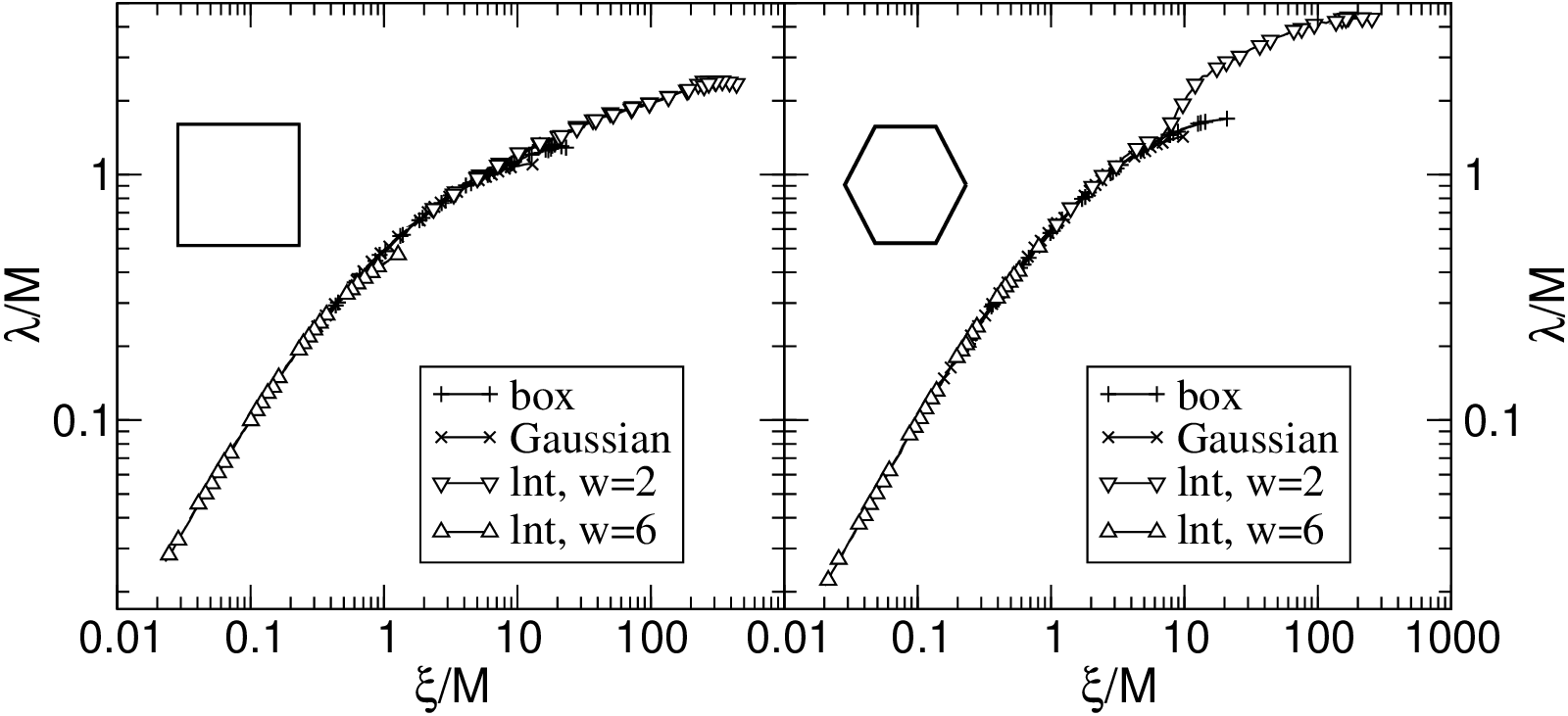}}
\caption{
  FSS plots of $\lambda_M(E,w)/M$ for square (left) and honeycomb
  (right) lattices. Note that data for different disorder distributions
  can be scaled simultaneously, except for the numerically most
  challenging $\ln t$ data at $w=2$ as shown in the right panel. }
\label{fig-fss-sq}
\label{fig-fss-hon}
\end{vchfigure}
We also calculate error estimates for $\xi$ taking into account the
accuracy of the raw localization lengths $\lambda$. Namely, we repeat
the FSS procedure several hundred times changing randomly the input
localization lengths $\lambda$ within its $1\%$ accuracy (of a Gaussian
distribution). Then we calculate the standard deviation of the obtained
values of the scaling parameter.  The errors --- as shown in the figures
below --- typically increase close to $E=0$ and are larger when larger
strip widths are used in FSS.

\subsection{Divergence of the localization lengths and further finite-size 
effects}
\label{sec-finite}

To investigate the divergence of the infinite-size localization lengths
at the band center we plot the scaling parameter $\xi$ in a doubly
logarithmic plot \cite{ISLOC}.  The deviation of the divergence from the
power-law behavior should be then easily seen. In most cases we observe
that at energies close to $E=0$ the divergence is slower than described
by a power-law.  Two examples are shown in Fig.\ \ref{fig-fss-change}.
\begin{vchfigure}[htb]
\centerline{\includegraphics[width=.95\textwidth]{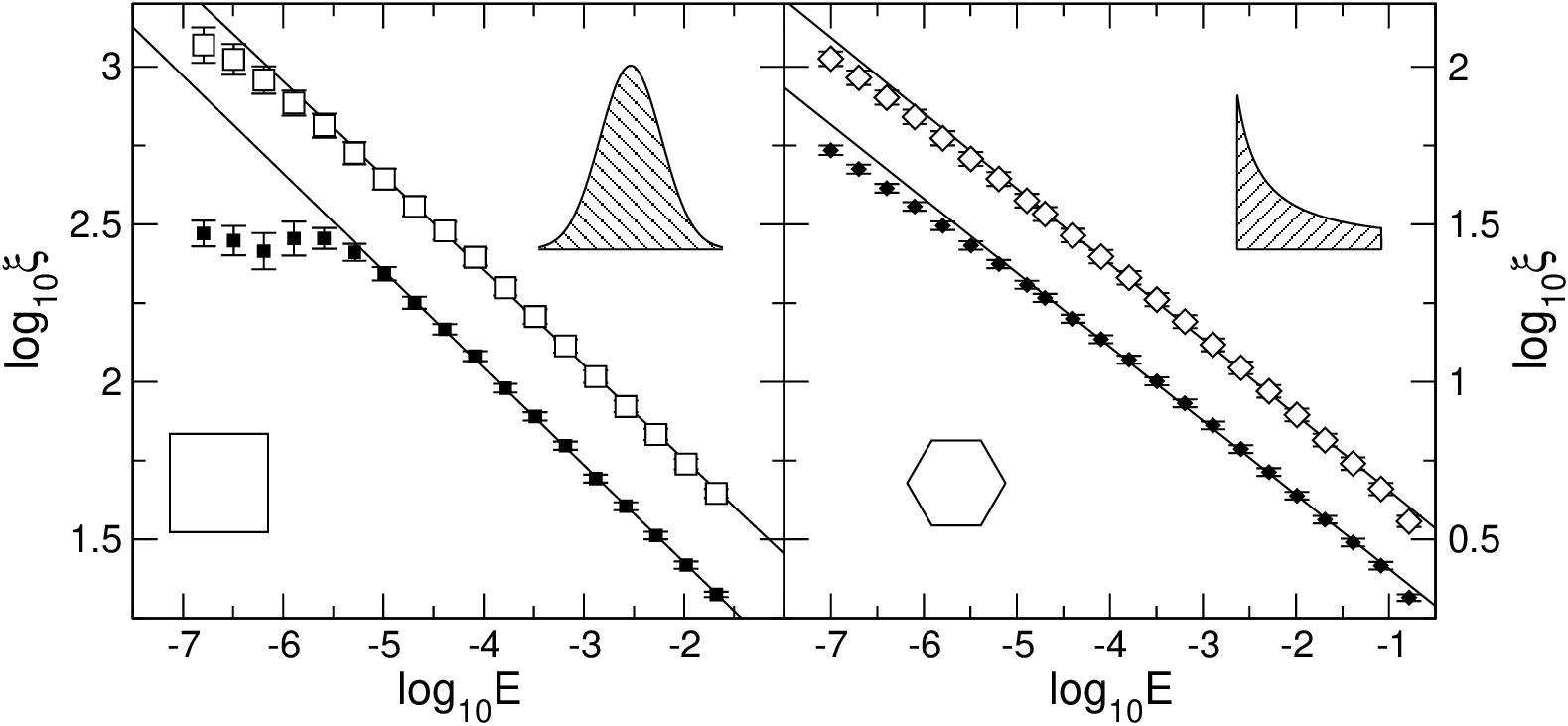}}
\caption{
  Scaling parameter $\xi$ as function of energy for a square lattice and
  Gaussian disorder distribution (left) and a honeycomb lattice with
  $\ln t$ disorder distribution at $w=6$ (right).  The small, filled
  symbols denote results for system widths $M=50$--$100$ and have been
  shifted downwards by $\log_{10} 0.25$ (left panel) and $\log_{10} 0.2$
  (right panel) for clarity. The large, open symbols indicate results
  for $M=110$--$160$ (Gaussian distribution) or $M=100$--$150$ ($\ln t$
  distribution).  }
\label{fig-fss-change}
\end{vchfigure}
The left panel presents a log-log plot of the energy dependence of the
scaling parameter $\xi$ for a Gaussian $t$ distribution with $c=0$ on a
square lattice. The curve obtained for smaller strip widths $M=50-100$
exhibits clear deviations from the straight line for smaller energies.
However, for larger widths ($M=110-160$) this deviation gets smaller and
the dependence is power-law in a wider energy range. Generally, in the
energy range where the $\lambda_{M}$ for small widths $M$ decreases
close to $E=0$ or diverges slower than for larger $M$, one can observe
that the values of the scaling parameter $\xi$ resulting from the FSS
are too small compared to the values obtained for larger systems, which
is manifested as a deviation from the power-law behavior of the
localization lengths. 

Therefore, the deviations from the power-law behavior as in Fig.\ 
\ref{fig-fss-change} appear to be finite-size effects that remain even
after FSS.  We performed such a check for all disorders and disorder
distributions.  It turns out that in almost all cases when the scaling
parameter exhibits a deviation from the power-law, this deviation is
smaller for larger system sizes, thus may be attributed to finite-size.
The only exception is the $\ln t$ distribution for $w=6$. An example for
a hexagonal lattice is shown in the right panel of Fig.\ 
\ref{fig-fss-change}.  In this case the results appear not to change
with the system size, thus the bending down of the line close to the
band center may reflect the change in the behavior of the localization
length, i.e., a crossover from power-law to the exponential form as in
Eq.\ (\ref{eq-power-law}).

\subsection{Numerical problems for off-diagonal disorder with small $t$ and
  $\ln t$ distribution}
\label{sec-hopping}

From the TMM equation (\ref{eq-tmm}) it follows that the division by
hopping elements $t^{||}_{n+1}$ is necessary to calculate the wave
function in the next step of TMM.  This may be a source of potential
numerical problems if the hopping elements are very close to zero. In
the case of rectangular and Gaussian distribution we therefore applied a
cutoff for small $|t^{||}_{n+1}|$ values and checked that the results do
not depend on the cut-off.
Furthermore, even for hopping amplitudes larger than the cut-off, the
ratio $[t^{||}_{n+1}]^{-1}t^{||}_{n}$ in Eq.\ (\ref{eq-tmm}) may become
large and lead to a large value of the corresponding component of the
wave function. In the next steps this large value will grow even further
which might lead to the loss of numerical accuracy due to round-off
errors. The growth of the wave vector component is normally suppressed
by the repeated reorthonormalization of the wave functions.  In our
calculations we performed the reorthonormalization after a fixed number
of 10 TMM steps.  We checked that in the case of rectangular and
Gaussian disorder distribution this is sufficient to guarantee numerical
stability. Note that an automatic reorthogonalization scheme as is
customarily used for diagonal disorder does not work so well in the
present case of off-diagonal disorder \cite{MilRSU00}.

One might expect that similar problems do not appear in the case of the
$\ln t$ distribution as the elements are never zero by definition.
However, in the $\ln t$ distribution the most probable hopping elements
are small elements $\gtrsim \exp(-w/2)$. Thus there are many large
$t$-ratios $\sim \exp(w/2)$ and wave function components. This has a
striking effect on the stability of the calculation as shown, e.g., in
Fig.\ \ref{fig-wdis}. 
\begin{vchfigure}[htb]
\centerline{\includegraphics[width=.95\textwidth]{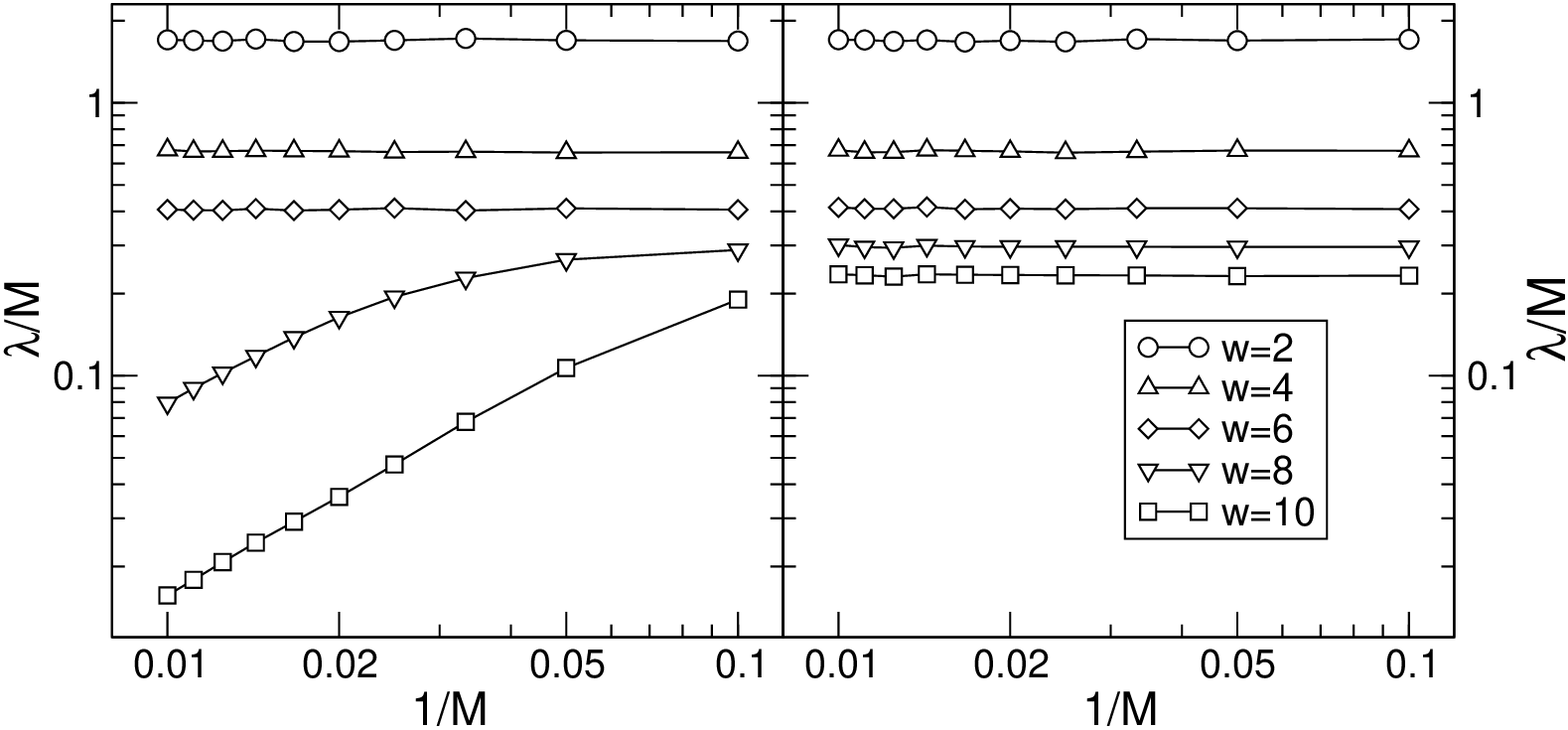}}
\caption{
  Reduced localization lengths $\lambda_M/M$ for square lattice and $\ln
  t$ distribution. In the left panel, reorthonormalization of the wave
  functions has been performed after every 10th TMM step. In the right
  panel, reorthonormalization is being done after each TMM step.  }
\label{fig-wdis}
\end{vchfigure}
When the number of TMM steps between renormalizations is kept at $10$,
the localization lengths seem to indicate localized behavior for
increasing $w$ values. However, renormalizing after each TMM step, we
see that there is no localization, rather, the states remain critical.
Let us emphasize that this effect appears for larger disorder where
naively one would expect numerical stability to become better (as for
diagonal disorder). Generally, all runs with $w>6$ need to be done with
renormalization after each TMM step.
It is also worth noting that, as the wave function may grow in one TMM
step up to $\exp(w)$, for sufficiently large $w$ numerical problems will
arise even if the wave functions are normalized after each TMM step.

In the present manuscript, we therefore calculate our localization
lengths for $w=6$ disorders using reorthonormalization after each $5$th
TMM step.

\subsection{Estimates of the divergence exponents}
\label{sec-final}

\begin{SCfigure}[6][htb]
\includegraphics[width=.45\textwidth]{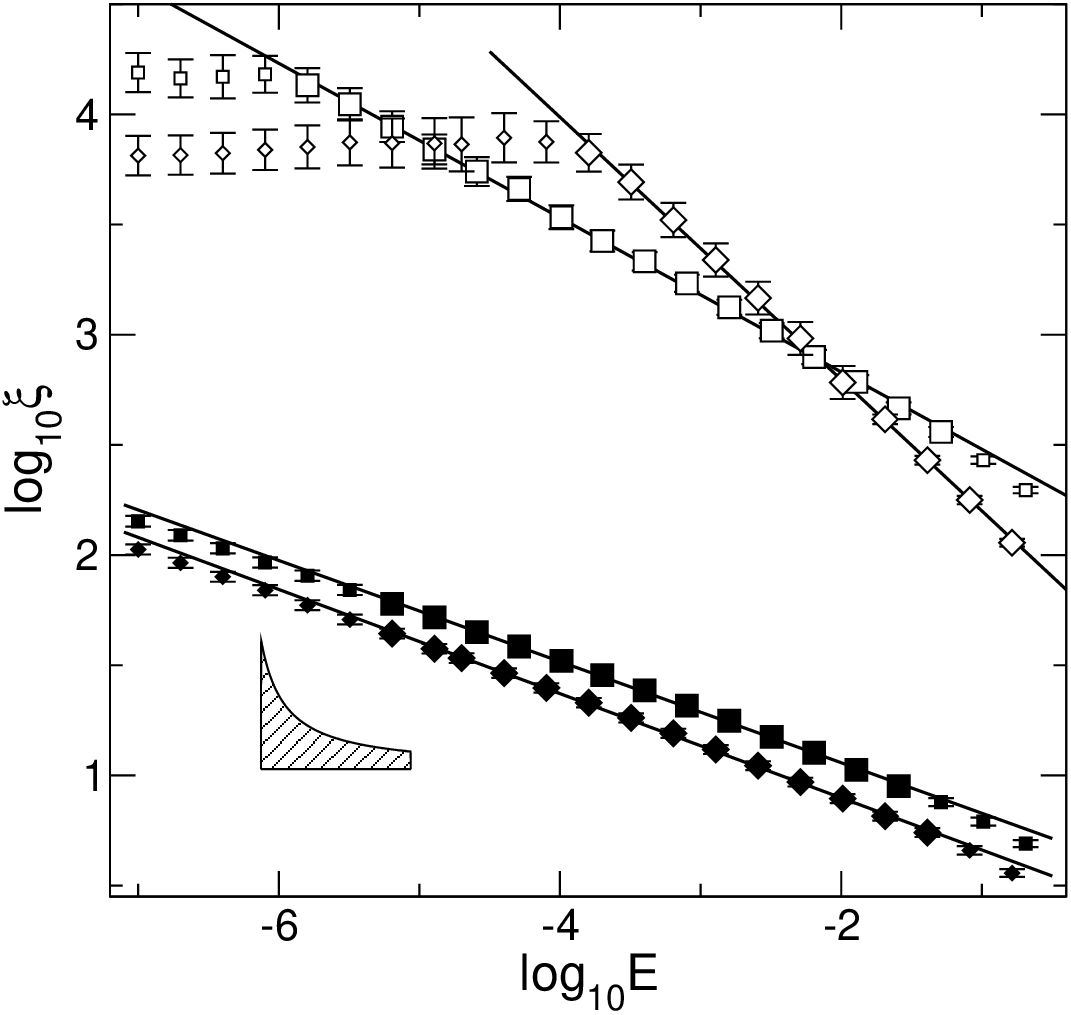}
\caption{
  Variation of the infinite-size localization length $\xi$ with $E$ for
  $\ln t$ disorder distribution with $w=2$ (open symbols) and $w=6$
  (filled symbols). Squares denote data for a square lattice, diamonds
  denote a honeycomb lattice.  Only the points marked by large symbols
  were used to fit the straight lines.  }
\label{fig-sc-lnt}
\end{SCfigure}

In Figs.\ \ref{fig-sc-box} and \ref{fig-sc-lnt} we show the dependence
of the scaling parameter $\xi$ versus the energy in the
double-logarithmic plot for rectangular, Gaussian and logarithmic $t$
distributions, respectively.
\begin{vchfigure}[htb]
\centerline{
\includegraphics[width=.45\textwidth]{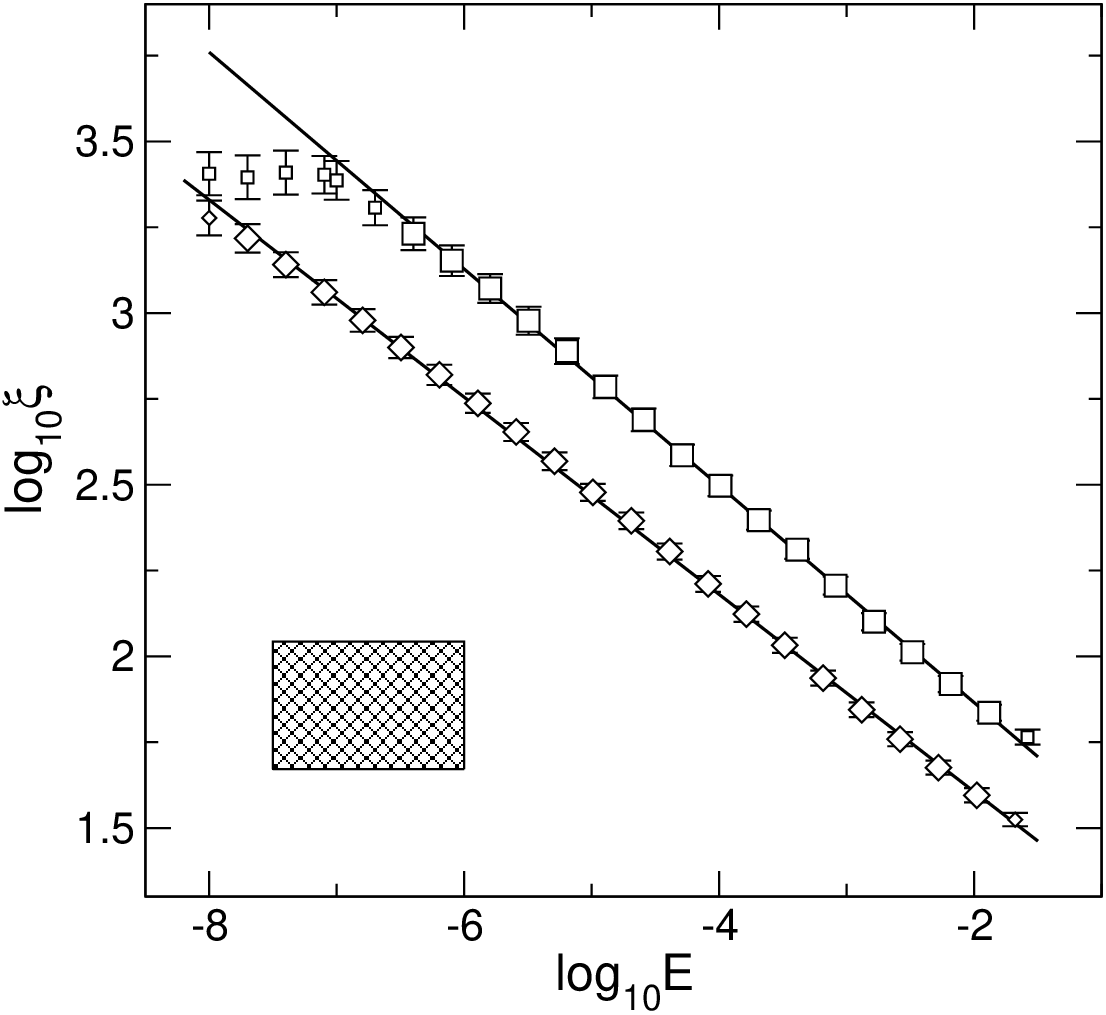}
\hfil
\includegraphics[width=.45\textwidth]{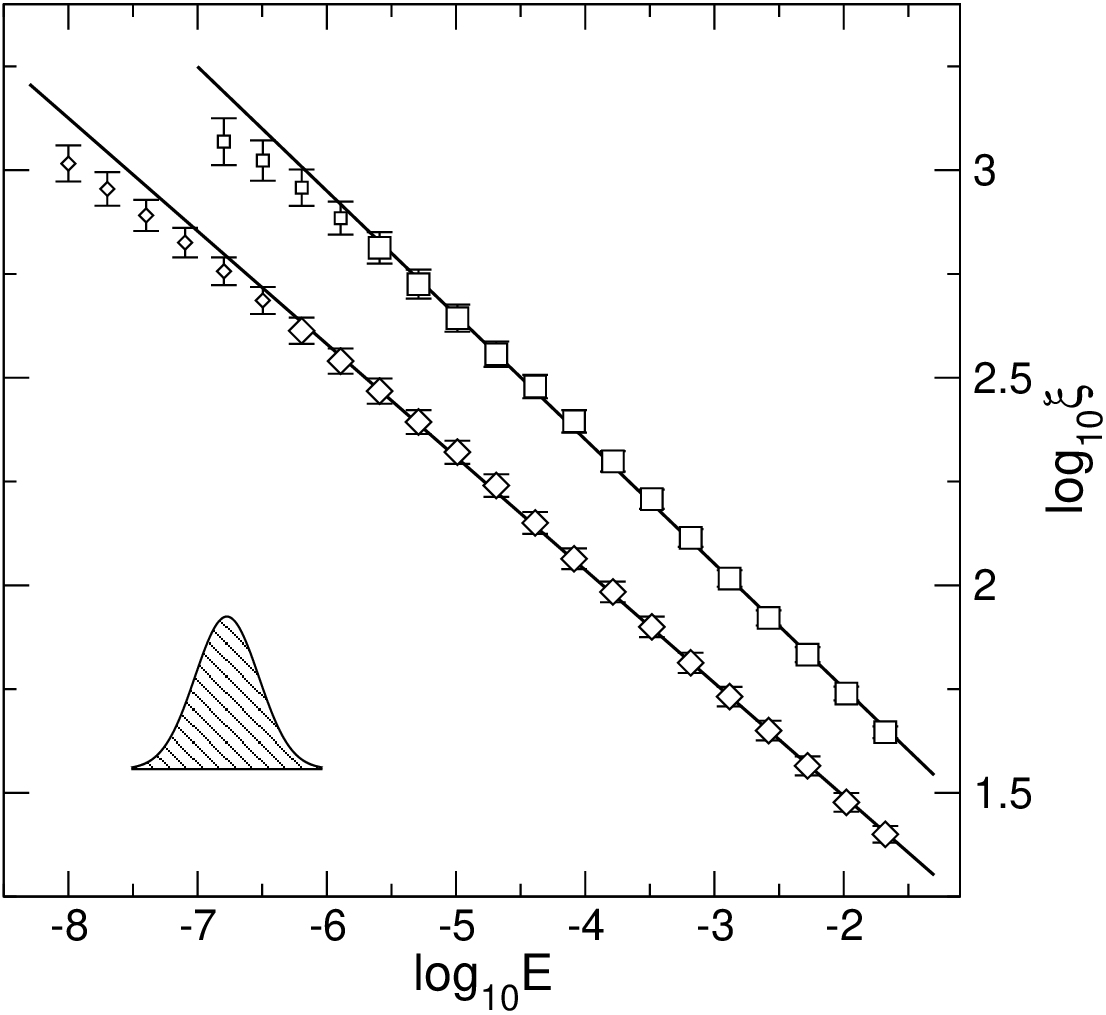}
}
\caption{
  Variation of the infinite-size localization length $\xi$ with $E$ for
  rectangular (left) and Gaussian (right) disorder distribution. Squares
  denote data for a square lattice, diamonds denote a honeycomb lattice.
  Only the points marked by large symbols were used to fit the straight
  lines.
}
\label{fig-sc-box}
\label{fig-sc-gauss}
\end{vchfigure}
The corresponding exponents are collected in Table \ref{tab-res}. They
are all in the range of $0.2-0.6$.  The values depend on the disorder
distribution and parameters, generally the exponents are larger for
weaker disorders when the localization lengths are larger.  The
differences between square and honeycomb lattices appear smaller than
the error for most distributions except for the $\ln t$ distribution
with $w=2$. In that case the exponent is almost $2$ times larger on the
honeycomb lattice. Note that this is precisely the case when the DOS do
not exhibit a peak at the band center. Thus the large change in the
exponent may be related to the low density of states in the case of the
hexagonal lattice. Of course, this also makes the calculation of the
localization lengths more time consuming therefore for this system we
were able to obtain the results for relatively large energies only.

We emphasize that the exponents do not fullfill the Chayes criterion
$\nu \geq 2/d$ \cite{ChaCFS86} with $d$ the lattice dimensionality.
However, this is also not to be expected since this is not a true phase
transition and for small energies, we expect the crossover as indicated
in Eq.\ (\ref{eq-power-law}).

\section{The non-bipartite triangular lattice}
\label{sec-non-bipartite}

As shown in Sec.\ \ref{sec-bipartite}, the localization lengths $\lambda_M$
for even and odd system sizes are very similar in the limit of large
system widths. Apriori this not obvious since the odd-sized systems are
not strictly bipartite due to the periodic boundary conditions.  In
\cite{FabC00} it was shown that the state at the band center is expected
to be critical only for bipartite lattices.  The breakdown of
bipartitness leads to complete localization.  In our case, however, the
non-bipartitness is only at the boundaries, therefore we find that it is
negligible for large enough widths.

\begin{SCfigure}[6][htb]
\includegraphics[width=.35\textwidth]{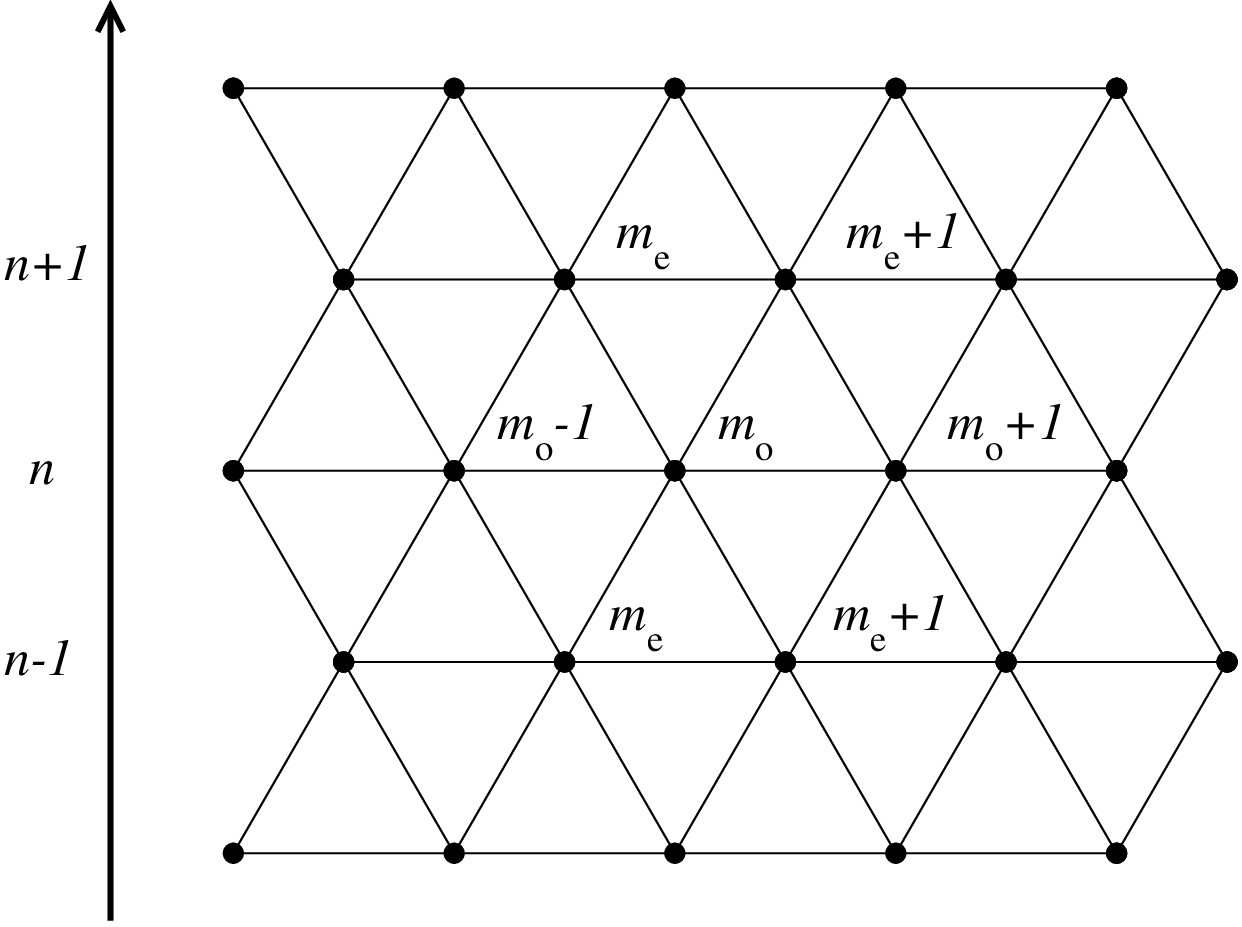}
\caption{
  The two-dimensional triangular lattice. The arrow shows the direction
  in which the TMM proceeds. The dots indicate lattice site and the
  solid lines show the connectivity of the lattice. The labelling of
  each sites in odd and even $n$ slices is indicated.}
\label{fig-triang}
\end{SCfigure}

Let us now turn to an example of a strictly non-bipartite lattice with
off-diagonal disorder such as the triangular lattice shown in Fig.\ 
\ref{fig-triang}.
In the triangular lattice each node is connected to the next and
previous slice by two connections instead of one \cite{SchO91}. The TMM
equations for this system are similar to Eq.\ (\ref{eq-tmm}) with the
exception that the matrices $t^{||}_n$ are no longer diagonal. Hence
our TMM procedure is essentially the same as for the triangular lattice
with diagonal disorder \cite{SchO91}, except that the connectivity
matrices $t^{||}_n$ will now be different for each slice. As before for
each TMM step the inversion of the $M \times M$ connectivity matrix
$t^{||}_n$ must be computed --- in practice we solve the equivalent set
of linear equations --- which significantly increases the computational
effort.

We performe the TMM calculations at the energy $E \rightarrow 0$ for
system sizes $M=10, 20, \ldots, 110$ and the centers of the rectangular
disorder distribution of hopping elements $c=0,0.1, \ldots, 0.5$. The
states at the band center are localized for all $c$ values in agreement
with \cite{FabC00}. The FSS plot for the data is shown in Fig.\ 
\ref{fig-fss-triang}, where the localized behavior of the states is
clearly visible.
\begin{vchfigure}[htb]
\centerline{\includegraphics[width=.95\textwidth]{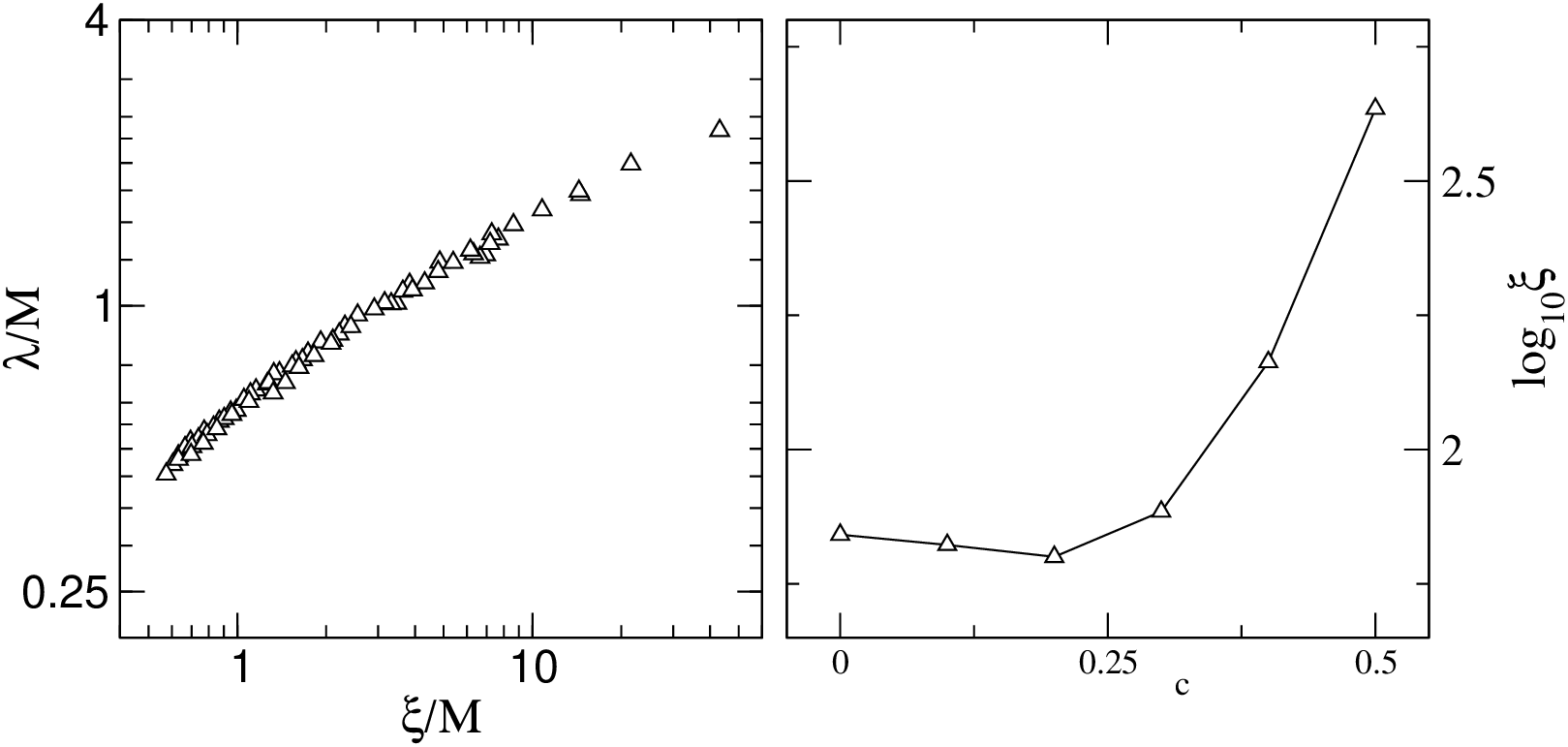}}
\caption{
  Left: Scaling plot for the triangular lattice with box disorder at
  E=$0$, $c=0, 0.1, \ldots, 0.5$ and system sizes $M=10, 20, \ldots,
  100$.  Right: dependence of the scaling parameter $\xi$ on the box
  distribution center $c$. }
\label{fig-fss-triang}
\end{vchfigure}
The values of the scaling parameters are shown in the inset; it may be
noted that the strongest localization (strongest disorder) appears for
$c=0.2$ which is a typical behavior for the chosen rectangular disorder
distribution \cite{EilRS98a}.

\section{Conclusions}
\label{sec-concl}

We calculated the localization lengths for different disorder
distributions on square and honeycomb lattices. We determined the energy
ranges in which the divergence of the localization lengths at the band
center is described by a power-law. We also computed the divergence
exponents which fall in the range $0.2-0.6$ and seem to depend on the
disorder parameters and in some cases on the lattice topology. 

For smaller energies we observe deviations from the power-law. However,
for most systems these deviations become smaller for larger system
widths $M$ and it appears that they are due to finite-size effects that
are retained even after FSS. The exception is the $\ln t$ distribution
with $w=6$, where below an energy $E \approx 10^{-5}$ we observe a
size-independent deviation from the power-law. This may be an indication
of the crossover to the non-power-law behavior such as predicted in Ref.
\cite{FabC00} or to a power-law with a different exponent.  However,
there is also a possibility that this may be an still an effect of as of
yet unknown further numerical problems which we already encountered for
strong $\ln t$ disorder.

\begin{acknowledgement}
  It is a pleasure to thank George Rowlands for stimulating discussions.

  This article is dedicated to Michael Schreiber on the occasion of his
  50th birthday. We are grateful to him for much stimulating material
  and encouraging structure over the last decade. 
\end{acknowledgement}

%
%


\begin{thebibliography}{10}
\bibitem{ISLOC} The FSS procedure of Ref.\ \protect\cite{MacK83}
  estimates the infinite-size localization lengths from the scaling
  parameter $\xi$ by extending the calculation to the strongly localized
  regime where $\lambda_M/M \approx \xi/M$. Here we are not in this
  regime and thus our $\xi$ are only proportional to the true
  infinite-size localizaton lengths. However, this is irrelevant when
  comparing with Eqs.\ (\protect\ref{eq-power-law}).
\end{thebibliography}

\end{document}